\documentclass[aps,prl,preprint,superscriptaddress, tightenlines, floatfix]{revtex4}

\usepackage{graphics}
\usepackage{longtable}
\usepackage{dcolumn}
\usepackage{amsmath}

\newcommand{\isotope}[3]{{}_{#3}^{#2}\text{{#1}}}
\newcommand{\sm}{\isotope{Sm}{152}{}}
\newcommand{\gd}{\isotope{Gd}{154}{}}

\newcommand{\pb}{\isotope{Pb}{208}{}}

\newcommand{\git}{School of Physics, Georgia Institute of Technology, Atlanta,
Georgia 30332-0430, USA}

\newcommand{\rochester}{Nuclear Structure Research Laboratory, Department of
Physics, University of Rochester, Rochester, New York, 14627, USA}
\newcommand{\guelph}{Department of Physics, University of Guelph, Guelph,
Ontario N0B 1S0, Canada}
\newcommand{\triumf}{TRIUMF, 4004 Wesbrook Mall, Vancouver, British Columbia
V6T 2A3, Canada}
\newcommand{\ukyp}{Department of Physics and Astronomy, University of Kentucky,
Lexington, Kentucky  40506, USA}
\newcommand{\ukyc}{Department of Chemistry, University of Kentucky,
Lexington, Kentucky  40506, USA}
\newcommand{\ncsu}{Department of Physics, North Carolina State University, Raleigh, North Carolina, 27695-8202, USA}
\newcommand{\llnl}{Lawrence Livermore National Laboratory, Livermore, California 94551, USA}

\begin{document}


\title{Shape Coexistence and Mixing in $^{152}$Sm}


\author{W.~D.~Kulp}
\author{J.~L.~Wood}
\affiliation{\git}
\author{P.~E.~Garrett}
\affiliation{\guelph}
\affiliation{\triumf}
\author{C.~Y.~Wu}
\altaffiliation[Present address:  ]{\llnl}
\affiliation{\rochester}
\author{D.~Cline}
\affiliation{\rochester}
\author{J.~M.~Allmond}
\affiliation{\git}
\author{D.~Bandyopadhyay}
\altaffiliation[Present address:  ]{\triumf}
\affiliation{\ukyp}
\author{D.~Dashdorj}
\affiliation{\ncsu}
\author{S.~N.~Choudry}
\affiliation{\ukyp}
\author{A.~B.~Hayes}
\affiliation{\rochester}
\author{H.~Hua}
\affiliation{\rochester}
\author{S.~R.~Lesher}
\altaffiliation[Present address:  ]{\llnl}
\affiliation{\ukyp}
\author{M.~Mynk}
\affiliation{\ukyc}
\author{M.~T.~McEllistrem}
\affiliation{\ukyp}
\author{C.~J.~McKay}
\affiliation{\ukyp}
\author{J.~N.~Orce}
\affiliation{\ukyp}
\author{R.~Teng}
\affiliation{\rochester}
\author{S.~W.~Yates}
\affiliation{\ukyp}


\date{\today}

\begin{abstract}
  Experimental studies of $\sm$ using multiple-step Coulomb excitation and inelastic neutron scattering provide key data that clarify the low-energy collective structure of this nucleus.
  No candidates for two-phonon $\beta$-vibrational states are found.
  Experimental level energies of the ground-state and first excited ($0^+$ state) rotational bands, electric monopole transition rates, reduced quadrupole transition rates, and the isomer shift of the first excited $2^+$ state are all described within $\sim$10$\%$ precision using two-band mixing calculations.
  The basic collective structure of $\sm$ is described using strong mixing of near-degenerate coexisting quasi-rotational bands with different deformations.
\end{abstract}


\maketitle


  Low-energy collective structure in nuclei is a fundamental manifestation of simple behavior in finite many-body quantum systems.  
  Nuclear collectivity is divided into two basic types:  rotational and vibrational \cite{Bohr1975b}. 
  Vibrational behavior is suggested to be dominant in spherical nuclei and rotational behavior in deformed nuclei.  
  Deformed nuclei also are suggested to be capable of vibrations about an equilibrium deformed shape.  
  The emergence of predominantly prolate spheroidal shape moments suggested two vibrational modes in deformed nuclei:  ``gamma'' vibrations ($Y_{22} + Y_{2-2}$ multipole mode) and ``beta'' vibrations ($Y_{20}$ multipole mode).  
  There is a voluminous literature that discusses one-phonon $\beta$-vibrational and $\gamma$-vibrational states in deformed nuclei and the lowest-lying excited $K^\pi = 0^+$ and $2^+$ states, respectively, are generally identified with these modes.  
  However, vibrations in quantum systems should exhibit multi-phonon eigenstates.  
  Evidence for multiple (two) phonon excitations in deformed nuclei is sparse and has been difficult to obtain.  
  The best examples are limited to evidence for two-phonon $\gamma$ vibrations, but controversy over this structural interpretation exists (see, e.g., \cite{Garrett1997, Fahlander1996, Hartlein1998, Oshima1995, Borner1991, Wu1994, Burke1994}).
  There is no unequivocal evidence for two-phonon $\beta$ vibrations.  

  The nucleus $\sm$ is particularly well-suited as a case study for the existence of multi-phonon $\beta$ and $\gamma$ vibrations in a deformed nucleus.  
  It has one of the lowest-energy candidate $\beta$ vibrations in any deformed nucleus (and a fairly low-energy candidate $\gamma$ vibration), such that 2- and even 3-phonon excitations should be below the pairing gap, if they exist.
  Indeed, recently it has been suggested \cite{Garrett2001} that $\sm$ (and its neighboring isotone, $\gd$) are the best candidates for establishing the $\beta$-vibrational mode in deformed nuclei.  
  This suggestion is supported by exploration of phase transitions in nuclei which places $\sm$ right at a critical point between spherical and deformed nuclear phases and  supports the expectation of multi-phonon vibrational behavior \cite{Casten2001a}.

  To address the expectation that these simple quantum excitation modes should exist in $\sm$, we have carried out a very high statistics study using multiple-step Coulomb excitation (multi-Coulex).  
  This study was made using the Gammasphere array \cite{Lee1996}  of Compton-suppressed Ge detectors in conjunction with the CHICO charged-particle detector array \cite{Simon2000}.  
  The experiment was conducted using a beam of $\sm$ ($E=652$~MeV, an energy insufficient to surmount the Coulomb barrier) incident on a thin $\pb$ target (400 $\mu$g/cm$^2$, 99.86$\%$ enrichment) at the Lawrence Berkeley National Laboratory's 88-Inch Cyclotron.  
  Signals from two ions detected by CHICO in coincidence with at least one ``clean'' $\gamma$ ray signal in Gammasphere (i.e., a p-p-$\gamma$ coincidence) triggered an event.  
  The CHICO array provided kinematic characterization of scattered ions and recoiling target nuclei so that Doppler corrections could be applied to the $\gamma$ rays emitted from the Coulomb-excited beam nuclei.  
  High angular resolution was provided both by CHICO, which has 1$^{\circ}$ angular resolution, and by Gammasphere, which was operated with 104 Ge detectors.  
  A total running time of 62 hours provided 7 $\times$ 10$^8$ p-p-$\gamma$, 8 $\times$ 10$^7$ p-p-$\gamma$-$\gamma$, and 1 $\times$ 10$^7$ p-p-$\gamma$-$\gamma$-$\gamma$ events.

\begin{figure}[htbp]
\resizebox{8.59 cm}{!}{\includegraphics*{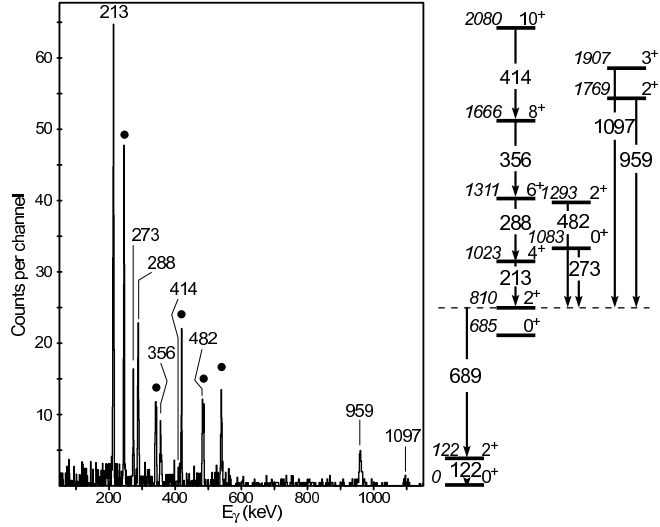}}
\caption{\label{689 gate} Double-coincidence gate indicating $\gamma$ rays in coincidence with the $2^+_2$ (810) $\rightarrow 2^+_1$ (122) $\rightarrow 0^+_1$ (0), 122 and 689 keV cascade $\gamma$-ray transitions in $\sm$.  Lines indicated by a dot above the transition are ``contaminants'' from an overlap with the $13^-_1$ (2833) $\rightarrow 12^+_1$ (2149) 685 - $2^+_1$ (122) $\rightarrow 0^+_1$ (0) 122~keV double-coincidence gate.}
\end{figure}

\begin{figure}[htbp]
\resizebox{8.59 cm}{!}{\includegraphics*{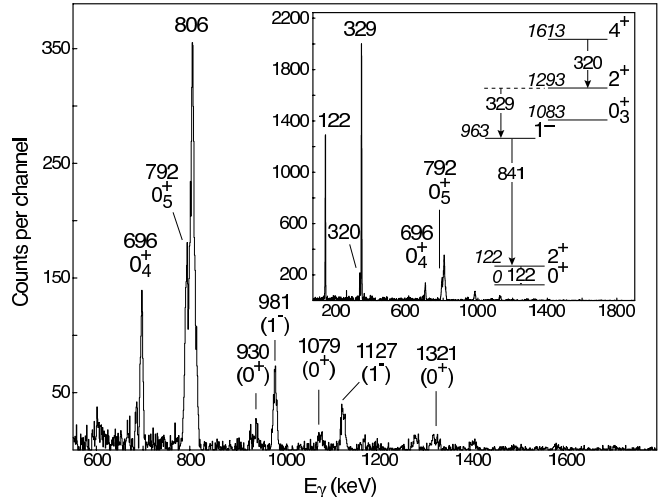}}
\caption{\label{841 gate} Coincidence gate on the $1^-_1$ (963) $\rightarrow 2^+_1$ (122) 841 keV $\gamma$-ray transition in $\sm$.  Details of the transitions are discussed in the text.}
\end{figure}

  The spectroscopic information pertinent to the existence of multi-phonon $\beta$-vibrational states is presented in Figs.~\ref{689 gate}  and \ref{841 gate} which show the $\gamma$ rays in coincidence with the 689-122~keV ($2^+_{2} \rightarrow 2^+_1 \rightarrow 0^+_1$) cascade and 841 keV ($1^-_{1} \rightarrow 2^+_1$) transitions, respectively.  
  Figure \ref{689 gate} reveals feeding of the $2^+_2$ 810~keV state (the ``$2^+_\beta$'' state):  
  there are strong $\gamma$ rays at 213, 288, 356, and 414~keV which correspond to the known \cite{Artna-Cohen1996} rotational band built on the $0^+_2$ 685~keV state;
  there are strong $\gamma$ rays at 273 and 482~keV which arise from the pairing isomeric band \cite{Kulp2005a} built on the $0^+_3$ 1083~keV state;
  and the strong lines marked with solid dots are transitions in the ground-state band that are due to 685-122 coincidences, where the 685~keV $\gamma$ ray is a known $13^-_1$ (2833) $\rightarrow 12^+_1$ (2149) transition \cite{Artna-Cohen1996}.
  At a much higher energy there are two $\gamma$ rays that correspond to a known transition of 959~keV from a $2^+$ state at 1769~keV \cite{Artna-Cohen1996} and a 1097~keV transition, which we propose as de-exciting a new state in $\sm$ at 1907~keV.
  
  The 841 gate shown in Fig.~\ref{841 gate} exhibits feeding of the $1^-_1$ 963~keV state and provides an important view of $0^+$, $1^-$, and $2^+$ states in $\sm$.
  The 320 and 329~keV $\gamma$ rays come from the pairing isomeric band (the 320 feeds through the 329~keV transition \cite{Kulp2005a});  the 696~keV $\gamma$ ray arises from the known \cite{Artna-Cohen1996} $0^+$ 1659~keV state; the 792~keV $\gamma$ ray comes from a new $0^+$ state at 1755~keV (see discussion later); the 806~keV $\gamma$ ray de-excites the $2^+$ 1769~keV state (see above); other $\gamma$ rays are all known (see below) to feed the 963~keV state and nearly all are observed in the $\beta$ decay of $^{152}$Pm ($T_{1/2} = 4.1$ m, $J^{\pi} = 1^+$) which preferentially populates states in $\sm$ with $J \le 2$ \cite{Artna-Cohen1996}.
  
  Clearly, the present multi-Coulex reaction achieved good population of $0^+$ states in $\sm$.
  However, there is no evident candidate $0^+$ state corresponding to a two-phonon $\beta$-vibration.
  A two-phonon to one-phonon transition is expected to have significant collective $E2$ strength (cf. Fig.~2 in \cite{Casten2001a}).
  The $0^+_2 (810) \rightarrow 2^+_1 (122)$ 563~keV transition is a collective transition with a $B(E2)$ of 33 Weisskopf units (W.u.) \cite{Artna-Cohen1996}.
  If the $0^+_2$ 685~keV state is a one-phonon vibrational state, then the transition to this level from a  two-phonon state should have a $B(E2) = 45$ W.u. \cite{Casten2001a}; our further investigation of this issue is described below.

  To complement the multi-Coulex results, we carried out an $(n,n'\gamma)$ study of $\sm$ at the University of Kentucky with monoenergetic neutrons.
  The data obtained included excitation functions ($E_n$ in 0.1 MeV increments from 1.2 to 3.0 MeV), $\gamma$-ray angular distributions, Doppler-shifted $\gamma$-ray energy profiles, and $\gamma-\gamma$ coincidences (details of similar analyses may be found in \cite{Belgya1996}).
  These data provide comprehensive information on low-spin states, including spins, decay branches, and lifetimes to $\ge 2.1$ MeV excitation.
  Here we focus on candidate states for a $0^+$ two-phonon $\beta$ vibration.
  
\begin{figure}
\resizebox{8.59 cm}{!}{\includegraphics*{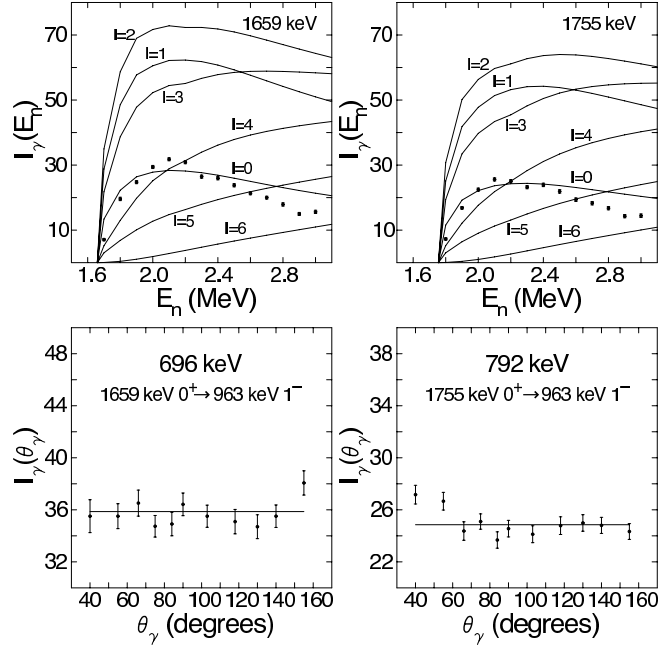}}
\caption{\label{excit} Gamma-ray excitation functions and angular distributions for transitions that depopulate the $0^+_4$ 1659 and $0^+_5$ 1755 keV levels, respectively.}
\end{figure}

\begin{figure}
\resizebox{8.59 cm}{!}{\includegraphics*{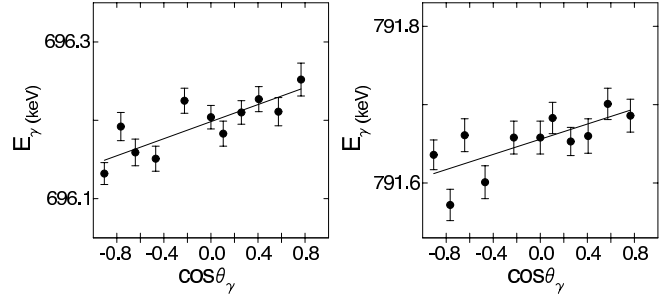}}
\caption{\label{lifetimes}  Doppler shift plots of the 696 and 792 keV $\gamma$ rays following the ($n,n'\gamma$) reaction on $\sm$.  Lifetimes of the $0^+_4$ 1659 and $0^+_5$ 1755 keV levels were extracted from these data.}
\end{figure}

  The previously known \cite{Artna-Cohen1996} excited $0^+$ states in $\sm$ are at  684.7 (the putative one-phonon $\beta$ vibration \cite{Garrett2001}), 1082.9 (a pairing isomer \cite{Kulp2005a}), and 1659.5 keV.
  We establish a new $0^+$ state at 1755.0~keV and tentatively assign $J^\pi=0^+$ or $1^-$ states at 1892.4, 1944.7, 2042.8, 2091.2, and 2284.9~keV based upon spectroscopy selection rules  (cf. Fig. \ref{841 gate}).
  We focus here on the 1659.5 and 1755.0~keV $0^+$ states.  
  Excitation function data for these states are shown in Fig.~\ref{excit}.
  These data can be compared with the plotted curves of theoretical direct population of the level as a function of neutron energy and level spin, calculated with the CINDY computer code \cite{CINDY1973} using input optical model parameters from the RIPL-2 database \cite{RIPL2}.
  The best agreement between plotted data and theoretical curves indicates the 1659.5 and 1755.0~keV levels are spin-0 states.
  From Doppler-shifted energies of decaying $\gamma$ rays, lifetimes of $177^{+65}_{-41}$ and $242^{+129}_{-66}$~fs are deduced, respectively, for the 1659.5 and 1755.0~keV states, and the data are shown in Fig \ref{lifetimes}.
  From the decay paths established (from the present work and \cite{Artna-Cohen1996}) for these states we deduce $B(E2; 0^+, 1659 \rightarrow 2^+_2, 810) = 5$ W.u. and $B(E2; 0^+, 1755 \rightarrow 2^+_2, 810) < 5$ W.u.  
  From these data we conclude that no candidates for two-phonon $\beta$ vibrations exist in $\sm$.
  Thus, it does not seem useful to interpret the $0^+$ 684.7~keV state as a $\beta$ vibration.
  We discuss the large $B(E2; 0^+_2 \rightarrow 2^+_1)$ below.
  (The collectivity implied by the observation of the 959 and 1097~keV $\gamma$ rays in Coulomb excitation, cf. Fig.~\ref{689 gate}, is assigned to a $K^{\pi} = 2^+$ collective excitation associated with the $0^+_2$ 684.7~keV state, and this will be discussed in a forthcoming publication.)

  The absence of any candidate two-phonon $\beta$-vibrational $0^+$ state in the energy range studied, i.e., up to $\sim$2750~keV (cf. Fig.~\ref{689 gate}) suggests that the $0^+$ state at 685~keV should not be interpreted as a $\beta$ vibration.  
  Rather, we explore a shape coexistence interpretation of the ground band and ``$\beta$'' band in $\sm$, and have carried out a simple two-band mixing calculation.  
  The band energies for the less-deformed structure were taken from \cite{Bhat2000, Reich1998} the ground-state band of $^{148}$Ce and for the more-deformed structure from $^{154}$Sm:  these are less-deformed and more-deformed nuclei adjacent to the region formed by $^{150}$Nd, $^{152}$Sm, $^{154}$Gd, and $^{156}$Dy.
  
  The key to fixing the relative energies of these two bands lies in the $\rho^2(E0)$ values \cite{Kibedi2005,Wood1999} between them which is a maximum for $J\ge 4$.
  If the monopole strength is taken to be given by the Eu isotope shift \cite{Dorschel1984} at $N=90$, i.e., $\delta \langle r^2 \rangle = 0.39$ fm$^2$, then using \cite{Wood1999}
\begin{equation}
\rho^2_J(E0) \cdot 10^3 =  \frac{Z^2}{R^4} \left[\Delta \langle r^2 \rangle \right]^2 \cdot 10^3 \; \alpha_J^2 \; \beta^2_J
\end{equation}
the two unmixed bands should be degenerate at $J \simeq 6$.
  This will produce maximal mixing ($\alpha_{6}^{2} = \beta_{6}^{2}=1/2$) and maximal  $\rho^2_{6\rightarrow 6}(E0) = 1/4 Z^2/R^4 [ \delta  \langle r^2 \rangle ]^2 = 87\times 10^{-3}$.
  With this prescription, a mixing strength of $V_J = 310$ keV for all $J$ values reproduces the energies of the lowest two $K=0^+$ bands very closely, as shown in Table \ref{comparison}.
  The $\rho^2(E0)$ values between the bands are given in Table \ref{comparison}.
  The isomer shift between the $0^+_1$ and $2^+_1$ states $\delta_{2_1 - 0_1}\langle r^2 \rangle/\langle R^2 \rangle = 4.6 \times 10^{-4}$ (calc.), cf. $5.3\,(6) \times 10^{-4}$ (expt. \cite{Wu1969}).
  The agreement supports this description.

\begingroup  
\squeezetable 
\begin{table}
\caption{\label{comparison} Comparison of experimental data for the lowest two $K=0$ bands in $\sm$ with  values calculated using a two-band mixing model.}
\begin{ruledtabular}
\begin{tabular}{ccccc}
 & \multicolumn{4}{c}{Level energy (keV)}\\
 &  \multicolumn{2}{c}{Expt. \cite{Artna-Cohen1996}} &  \multicolumn{2}{c}{Calc.} \\
  \multicolumn{1}{c}{$J^{\pi}$} & 
  \multicolumn{1}{c}{band 1} &  \multicolumn{1}{c}{band 2}  & 
  \multicolumn{1}{c}{band 1} &  \multicolumn{1}{c}{band 2} 
\\
  \hline
$0^+$ &        0 &    685 &         4 &    692 \\
$2^+$ &   122 &    810 &    140 &    797 \\
$4^+$ &   366 &  1023 &    393 & 1023 \\
$6^+$ &   706 &  1311 &    730 & 1350 \\
$8^+$ & 1125 &  1666 & 1131 & 1758 \\
\\
 & \multicolumn{4}{c}{$10^3 \times \rho^2(E0)$}\\
Transition &  \multicolumn{2}{c}{Expt. \cite{Kibedi2005,Wood1999}} &  \multicolumn{2}{c}{Calc.} \\
\hline
$0^+_2 \rightarrow 0^+_1$ & \multicolumn{2}{c}{$51\,(5)$}   &\multicolumn{2}{c}{ 72} \\
$2^+_2 \rightarrow 2^+_1$ & \multicolumn{2}{c}{$69\,(6)$}   & \multicolumn{2}{c}{77}  \\
$4^+_2 \rightarrow 4^+_1$ & \multicolumn{2}{c}{$88\,(14)$\footnotemark[1]} & \multicolumn{2}{c}{84} \\
$6^+_2 \rightarrow 6^+_1$ &  \multicolumn{2}{c}{ }                  & \multicolumn{2}{c}{87}  \\
$8^+_2 \rightarrow 8^+_1$ &  \multicolumn{2}{c}{  }                 & \multicolumn{2}{c}{85}  \\
\\
 & \multicolumn{4}{c}{$B(E2)$ (W.u.)}\\
Transition &  \multicolumn{2}{c}{Expt. \cite{Artna-Cohen1996, Kulp2007a}} &  \multicolumn{2}{c}{Calc.} \\
\hline
$2^+_1 \rightarrow 0^+_1$ &  \multicolumn{2}{c}{$144\,(3)$}    &   \multicolumn{2}{c}{133} \\
$4^+_1 \rightarrow 2^+_1$ &  \multicolumn{2}{c}{$209\,(7)$}    &   \multicolumn{2}{c}{197}   \\
$6^+_1 \rightarrow 4^+_1$ &  \multicolumn{2}{c}{$245\,(9)$}    &   \multicolumn{2}{c}{230} \\
$2^+_2 \rightarrow 0^+_2$ &  \multicolumn{2}{c}{$167\,(16)$}  &   \multicolumn{2}{c}{170}  \\
$4^+_2 \rightarrow 2^+_2$ &  \multicolumn{2}{c}{$255\,(45)$}  &   \multicolumn{2}{c}{233}  \\
$0^+_2 \rightarrow 2^+_1$ &  \multicolumn{2}{c}{$ 33\,(4)$}     &    \multicolumn{2}{c}{ 31}  \\
$2^+_2 \rightarrow 0^+_1$ &   \multicolumn{2}{c}{$0.96\,(9)$}  &  \multicolumn{2}{c}{1} \\
$2^+_2 \rightarrow 2^+_1$ &   \multicolumn{2}{c}{$5.8\,(5)$}    &  \multicolumn{2}{c}{5}   \\
$2^+_2 \rightarrow 4^+_1$ &    \multicolumn{2}{c}{$18\,(2)$}     &     \multicolumn{2}{c}{23}  \\
$4^+_2 \rightarrow 2^+_1$ &    \multicolumn{2}{c}{$0.75\,(13)$} &  \multicolumn{2}{c}{1}\\
$4^+_2 \rightarrow 4^+_1$ &    \multicolumn{2}{c}{$6.1\,(11)$}  &  \multicolumn{2}{c}{5}   \\
$4^+_2 \rightarrow 6^+_1$ &    \multicolumn{2}{c}{$16\,(5)$}     &  \multicolumn{2}{c}{21} \\         
\end{tabular}
\end{ruledtabular}
\footnotetext[1]{Calculated using data from \cite{Artna-Cohen1996} and \cite{Klug2000}.}
\end{table}
\endgroup
 
  The $B(E2)$ values between, and within, these bands can be simply described using Grodzins' rule \cite{Grodzins1962} relating $E(2^+_1)$ to $B(E2;0^+_1 \rightarrow 2^+_1)$ and rotor J-value dependence of $E2$ matrix elements.
  Some important resulting values are given in Table \ref{comparison}.
  The pattern of  $B(E2)$ values is closely reproduced.
  In particular, the excited $K=0$ band is correctly described as more collective \cite{Kulp2007a} than the ground band and $B(E2;0^+_{2} \rightarrow 2^+_1)$ is well reproduced by the mixing, i.e., the significant strength of this transition does not necessitate invoking $\beta$-vibrational character.

  The present results are in disagreement with the concept of a $\beta$ vibration and with a critical point interpretation of the low-energy collective structure of $\sm$.  
  (We note that certain features of the low-energy collective structure of $\sm$ were the original motivating factor for suggesting critical point behavior in nuclei \cite{Iachello1998}.)  
  However, all of the spectroscopic data are consistent with coexisting collective structures \cite{Wood1992} in $\sm$.  
  
  A possible explanation of the present result is that the shape coexistence is arising from the influence of the $Z=64$ subshell gap at $N=90$, which can give rise to the coexistence of different proton pair structures, similar to the familiar intruder structures associated with major shells \cite{Wood1992}.  
  An indication of this effect is implicit in the discussions of Burke \cite{Burke2002a} and Garrett \cite{Garrett2001} based on single-proton transfer population of the states in $\sm$ \cite{Hirning1977}.
  We note that in \cite{Hirning1977} it was found that the $(t,\alpha)$ population of the $2^+_1$ and $2^+_2$ states in $\sm$ was in the ratio $\sim$2:1 and the strength summed to that 
  expected for the ground-state configuration of the $^{153}$Eu target nucleus.
  This translates into a ratio of $\beta_2/\alpha_2 = 0.707$ for the mixing amplitudes of the $2^+$ states, which can be compared with our band-mixing calculation for which $\beta_2/\alpha_2 = 0.7069$.

  The present interpretation, based on strong mixing of coexisting quasi-rotational bands and proton-pair structures near $Z=64$, can be directly investigated through systematic studies of the $N=90$ isotones.
  There is also a systematic picture \cite{Wu2003} of band mixing at $N=60$ near $Z=40$ which may possess similarities.
  
\begin{acknowledgments}
  We wish to thank colleagues at the LBNL 88-Inch Cyclotron and the University of Kentucky monoenergetic neutron facility for their assistance in these experiments.  
  This work was supported in part by DOE grants/contracts
DE-FG02-96ER40958 (Ga Tech), DE-AC03-76SF00098 (LBNL),  and by NSF awards PHY-0244847 (Rochester) and PHY-0354656 (Kentucky).

\end{acknowledgments}


\newpage

\end{document}